\documentclass[twocolumn,showpacs,preprintnumbers,amsmath,amssymb]{revtex4}
\usepackage{epsfig}
\usepackage{amsmath}

\begin{document}

\title{The Path-Star Transformation and its Effects on Complex Networks}

\author{Luciano da Fontoura Costa}
\affiliation{Institute of Physics at S\~ao Carlos, University of
S\~ao Paulo, PO Box 369, S\~ao Carlos, S\~ao Paulo, 13560-970 Brazil}

\date{3rd Nov 2007}

\begin{abstract}
A good deal of the connectivity of complex networks can be
characterized in terms of their constituent paths and hubs.  For
instance, the Barab\'asi-Albert model is known to incorporate a
significative number of hubs and relatively short paths.  On the other
hand, the Watts-Strogatz model is underlain by a long path and almost
complete absence of hubs.  The present work investigates how the
topology of complex networks changes when a path is transformed into a
star (or, for long paths, a hub).  Such a transformation keeps the
number of nodes and does not increase the number of edges in the
network, but has potential for greatly changing the network topology.
Several interesting results are reported with respect to
Erdos-R\'enyi, Barab\'asi-Albert and Watts-Strogats models, including
the unexpected finding that the diameter and average shortest path
length of the former type of networks are little affected by the
path-star transformation.  In addition to providing insight about the
organization of complex networks, such transformations are also
potentially useful for improving specific aspects of the network
connectivity, e.g. average shortest path length as required for
expedite communication between nodes.
\end{abstract}

\maketitle

\vspace{0.5cm}
\emph{`There is only one corner of the universe you can
be certain of improving, and that is your own self.' (A. Huxley)}

\section{Introduction} 

The inherent flexibility that complex networks have for representing a
wide range of discrete structures and systems has given rise to one of
the most powerful and dynamic research
subjects~\cite{Albert_Barab:2002, Newman:2003, Boccaletti:2006,
Costa_surv:2007}.  While a vast number of real-world structures have
been found to exhibit the small world effect (even the uniformly
random Erdos-R\'enyi -- ER -- and the scale free Barab\'asi-Albert --
BA -- complex networks are characterized by small average shortest
path), particularly important real-world systems (e.g. protein
interaction, WWW and Internet) present scale free organization.
Several other types of networks, including those called geographical,
have been defined and are subject of current investigations
(e.g.~\cite{Costa_PhysA:2003, Kaiser_Hilgetag:2004}).

Each specific category of complex networks exhibit intrinsic
properties such as small average degree and high clustering
coefficient (small world networks), scale free distribution of node
degrees (power law networks), and large average shortest path
(geographical networks).  The uniformly random ER networks can be well
described in terms of their average node degree (most nodes have
degree similar to the average).  Because of the intrinsic simplicity
of the ER networks, this model has served as a reference for
characterizing the other models of networks as being relatively
\emph{complex}.

Since each type of complex network presents distinctive features, some
of them will result particularly efficient for specific tasks.  For
instance, small world networks are a natural choice for implementing
solutions where a relatively short path can be found between any pairs
of nodes, which is an important feature favoring fast communication
between pair of nodes.  On the other hand, geographical networks tend
to present high clustering coefficient, implying a good robustness to
edge attacks and large average shortest path length (such networks are
among the few structures which are not small world).  Several
interesting problems arise when one considers the relative efficiency
of networks.  For instance, one may want to design networks which
optimize or suit specific criteria.  Another interesting possibility
is, given a complex network, to try to modify it in order to improve
its fitness with respect to some imposed criteria.  One of the few
approaches in the complex network literature addressing the latter
subject has been described in~\cite{Costa_PRE:2004}, where the effects
of complementing the connectivity by considering several strategies
were assessed respectively to improvements in the network resilience.

\begin{figure}
  \vspace{0.3cm}
  \centerline{\includegraphics[width=0.9\linewidth]{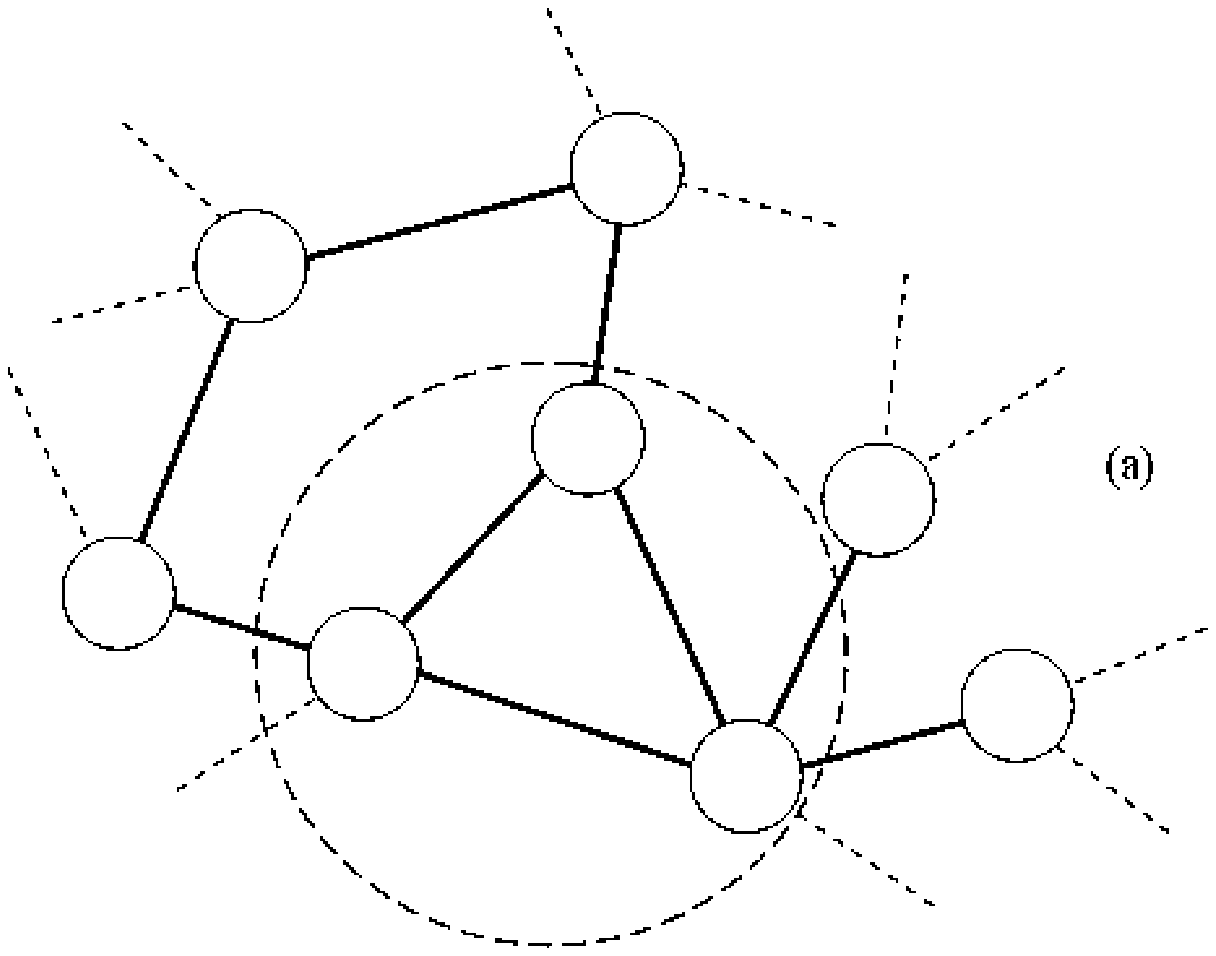}}
  \centerline{\includegraphics[width=0.9\linewidth]{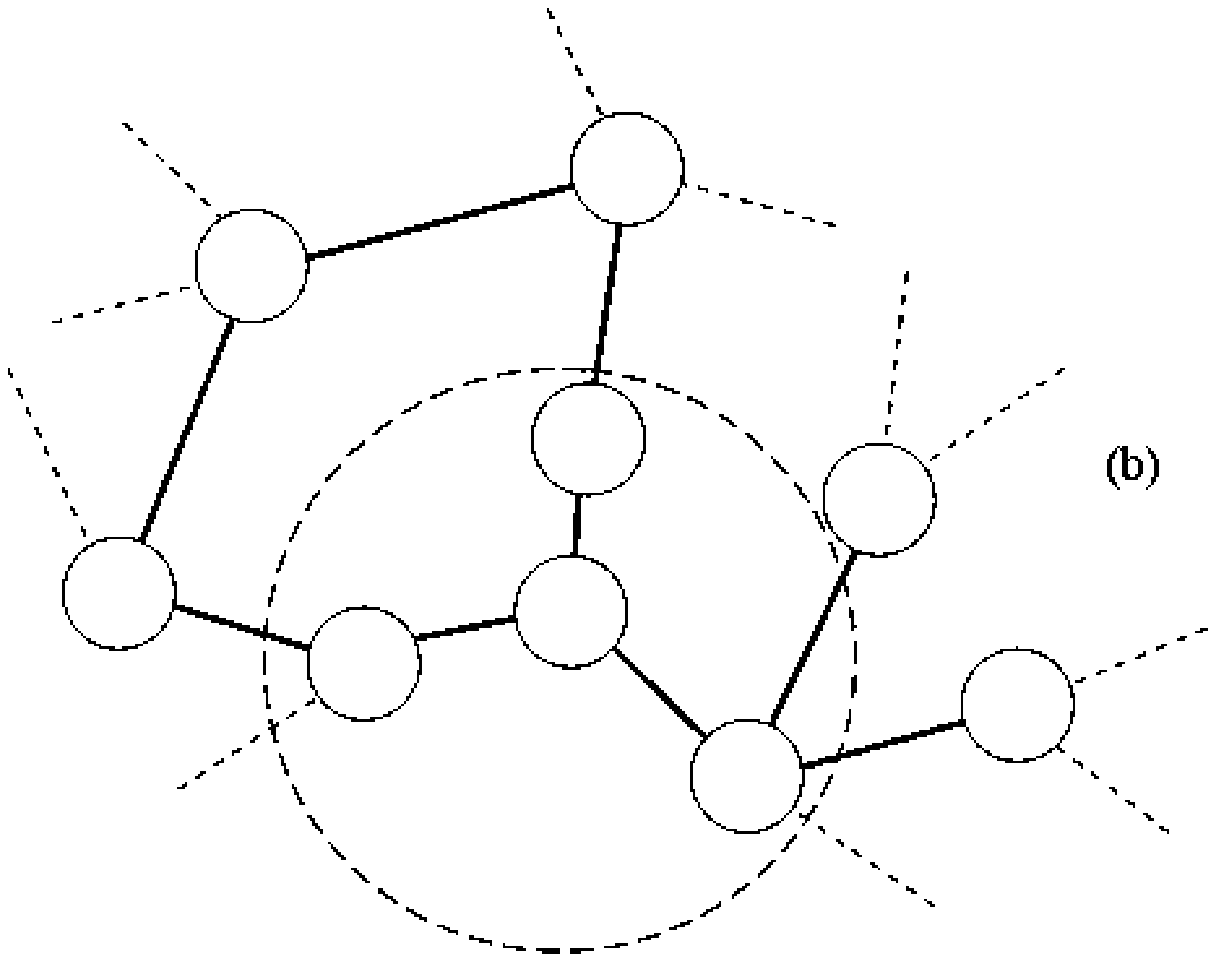}}
  \caption{Example of motif transformation: the triangle-star 
  conversion in a complex network: the triangle structure, shown within
  the circle in (a), is transformed into the encircled star in (b). 
  Note that such a conversion adds a new edge to the network.}
  \label{fig:tri_star}
\end{figure}

The modifications performed on a network, e.g. for improving some of
its specific features, can be divided into two main categories: (i)
\emph{global}, where the modifications are performed indiscriminately 
throughout the whole network; and (ii) \emph{local}, in which case the
modifications are restricted to specific parts of the network.
Examples of global modifications are provided by the rewiring schemes
which have been traditionally used in complex networks research
(e.g.~\cite{Milo:2003}), in which any connection may be changed.  An
example of local modification is to replace a motif~\cite{Milo:2002}
by another, e.g. a triangle by a star (see Figure~\ref{fig:tri_star}).
Observe that the triangle-star motif transformation is directly
related to the triangle-star conversion in electrical circuits and,
therefore, are particularly interesting for studying flow and random
walks in networks~\cite{Doyle_Snell:1984}. Such motif transformations
typically involve the same subset of network nodes.

Network modifications can also be characterized with respect to the
respectively implied costs.  So, we can define the following two main
categories: (a) \emph{costless}, involving no (or almost no) cost; and
(b) \emph{expensive}, in which case the modifications have a relevant
cost.  The cost is specific to each situation.  For instance, the
addition of nodes in the Internet will involve costs related to the
acquisition of new computers (servers).  Even modifications preserving
the number of edges and nodes (e.g. rewirings) may involve costs
related to the physical change of the connections, the period of time
the system will be down, as well as administrative expenses.

A critical research issue regards the effect of the modifications on
the network topological and dynamical characteristics.  For instance,
the rewiring scheme in~\cite{Milo:2003} changes several network
features, but does not affect the degree distribution.  Therefore, it
is interesting to characterize as completely as possible the effects
of network modifications.  This can be done by considering several
measurements~\cite{Costa_surv:2007}, especially those which are more
closely related to the expected features.  Interestingly, as
corroborated by the results reported in this work, the same type of
modification can have substantially different effects of the
topological features of the a network depending on the category of the
latter.

\begin{figure}
  \vspace{0.3cm}
  \centerline{\includegraphics[width=1.0\linewidth]{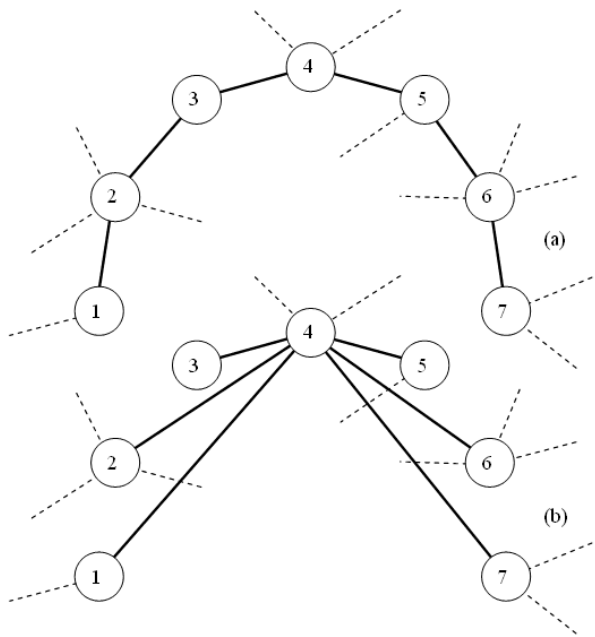}}
  \caption{The transformation of a path (a) into a star motif (b) can
  be obtained by replacing the edges along the path so that they
  connect the middle with the remainder nodes.  Observe that such a
  transformation keeps the number nodes and does not increase the
  number of edges, but has an important effect in decreasing most of
  the distances between the nodes. In the case where edges originally
  interconnect the nodes in the original path, the path-star
  transformation may also tend to increase the clustering
  coefficient.}  \label{fig:path_star}
\end{figure}

The present work investigates the effect of a special type of network
modification, especially with respect to improvements of specific
features of the network.  The considered modification is local and
involves replacing paths~\footnote{A \emph{path} in a network is a
succession of adjacent edges in which all the edges and all the
vertices are never repeated.  The length of the path corresponds to
its number of edges. (e.g.~\cite{Aldous:2003})} by stars, as
illustrated in Figure~\ref{fig:path_star}.  Such a type of motif
transformation can be understood as being related to the
generalization of the path star conversion in electrical
circuits. More specifically: a path is identified in the network; its
middle node is determined; and the connections among the nodes in the
path are replaced so that the middle node becomes directly connected
to all remainder nodes in the original path.  Observe that such a
path-star motif transformation~\footnote{Although paths are not
typically considered motifs, as in~\cite{Paulino:2007} here we extend
the concept of motifs to incorporate paths.} preserves the number of
nodes and does not increase the number of edges (note that eventual
edges already existing between the middle point and any of the other
nodes will become redundant), therefore implying low or no cost.  The
inverse motif transformation, namely taking a star into a path may
also be considered for improvement of complex networks.

The choice of the type of local transformation considered in this
work, among a large variety of alternative possible modifications
deserves some justification.  First, it implies a major alteration of
the topology of the network while preserving the number of nodes and
not increasing the number of edges.  Second, the two involved motifs
-- i.e. paths and stars -- are among the most important substructures
underlying the connectivity of networks.  Paths are also intrinsically
related to random walks on networks, while stars are related to hubs.
In addition, they are also very different one another and can even be
considered dual.  Paths have an intrinsic sequential, linear nature.
Contrariwise, stars are centralized motifs.  Also, the path-star (as
well as its inverse) preserves the overall connectivity of the overall
network, in the sense of not producing additional connected
components.  The effects of the path-star transformation (e.g. for
reducing the average distance between pair of nodes) depends on the
connectivity of the nodes involved in the chosen path and on the type
of network.  In the simplest case where the original network involves
only the path, the effects of the path-star transformation is clear
and include: (i) the average shortest path length (as well as the
network diameter) is reduced; (ii) the node degree distribution
changes, giving rise to a hub; and (iii) the clustering coefficient
remains zero.  Interestingly, the middle node remains with the higher
betweeness centrality.  Figure~\ref{fig:path_star} illustrates a more
generic path-star transformation. Observe that, unlike the
triangle-star conversion, this transformation does not increase the
number of edges because the middle point is used as center of the
star. Also, in case of long paths, the path-star transformation will
give rise to a hub.  Such a duality between long paths and hubs can
even be related to the justification of the presence of the latter in
some types of networks (i.e. hubs would be the result of the path-star
transformation required for optimization of specific network
features).

Although the effects of the path-star motif transformation can vary
with the specific chosen path and type of network, it is expected that
-- when applied to networks involving relatively long paths, this
transformation will tend to reduce the average shortest path length
and give rise to hubs.  The characterization of the effect of this
transformation provides one of the main motivations for the present
article.  The inverse motif transformation, namely star-path is
expected to have opposite effects.

This article starts by presenting the basic concepts and methods and
follows by presenting and discussing the results from the perspectives
of the average and standard deviation of the diameter, clustering
coefficient and average shortest path length as well as correlations
between such measurements and the length of the chosen path.

\section{Basic Concepts and Methodology}

This work focuses undirected networks, which can be represented by
symmetric adjacency matrices $K$, such that each existing edge $(i,j)$
implies $K(i,j)=K(j,i)=1$, with $K(i,j)=K(j,i)=0$ indicating absence
of the respective edge.  Each network contains $N$ nodes and $E$
edges.  The \emph{degree} of a node $i$, expressed as $k_i$, is equal
to the number of edges attached to that node.  The average node degree
considering the whole network is henceforth expressed as $\left< k
\right>$. The \emph{immediate neighbors} of a node $i$ are those nodes 
which are directly connected to $i$. The \emph{clustering coefficient}
of a node $i$, henceforth represented as $cc_i$ , is given by the
ratio between the total number of existing edges between the immediate
neighbors of $i$ and the maximum number of possible edges between
those nodes. The average clustering coefficient considering the whole
network is henceforth expressed as $\left< cc \right>$. Two edges are
\emph{adjacent} whenever they share one of their extremities.  A
sequence of adjacent edges form a \emph{walk}, whose length
corresponds to the number of involved edges.  A \emph{path} is a walk
where the edges and nodes never repeat themselves.  The \emph{shortest
path} between two nodes correspond to one (or more) of the paths
between those two nodes which involve the smallest possible number of
edges.  The average shortest path length considering the whole network
is henceforth expressed as $\left< spl \right>$. The \emph{diameter}
of the network, abbreviated here as $diam$, corresponds to the length
of the longest shortest path between any pair of nodes in the network.

This work considers the effect of the path-star transformation in
three principal types of networks: ER, BA and WS.  The BA networks
were grown by starting with $m0$ nodes and incorporating new nodes
with $m$ edges and preferential attachment
(e.g.~\cite{Albert_Barab:2002}), so that the average node degree is
given as $\left< k\right>=2m$.  The ER networks were obtained by
implementing connections between each pair of distinct nodes with
fixed probability.  The WS networks were obtained by starting with a
cycle of $N$ nodes and rewiring $20\%$ of the edges.  The parameters
of the models were chosen so as to yield similar node degrees for all
three models (e.g.~\cite{Costa_trails:2007, Paulino:2007}).

The experiments reported in this work involved a single path-star
transformation performed on a randomly chosen path.  More
specifically, a node $i$ of the network is randomly selected and a
random walk performed until no more unvisited nodes are available.
Another similar random walk is then performed starting again from $i$,
in order to explore the other extremity of the path.  The
so-obtained path is then transformed into a star by removing the edges
along the path and placing then between the middle node and all
remainder nodes.  Observe that the statistical relevance of the
adopted experimental procedure increases with the length of the
selected paths.

The effects of the path-star transformation in changing the network
topology can be quantified in terms of ratios between measurements
taken after and before the transformation.  The present work considers
the following ratios: 

\begin{eqnarray}
  R(diam) = diam(after) / diam(before) \\
  R(cc) = \left< cc \right>(after) / \left< cc \right>(before) \\
  R(spl) = \left< spl \right>(after) / \left< spl \right>(before) 
\end{eqnarray}

\section{Results and Discussion}

A total of 50 simulations were performed for each networks of type ER,
BA and WS models with respect to each of the four following
configurations: (a) $N=100$ and $m=3$; (b) $N=100$ and $m=5$; (c)
$N=200$ and $m=3$ and (d) $N=200$ and $m=5$.  The obtained results are
presented and discussed in the two following subsections with respect
to their average/standard deviation and correlations with the initial
path length.

\subsection{Average and Standard Deviation of Ratios}

The average and standard deviation of the ratios obtained for the
changes in diameter, clustering coefficient, average shortest path
length and length of the initial chosen path are given in
Table~\ref{tab:res} with respect to each of the above four
configurations and the three types of considered network models.

Table~\ref{tab:res} also shows, in its last group, the values of the
lengths of the randomly chosen paths used for the path-star
transformation.  Such lengths tended to present a relatively small
dispersion, especially for the cases involving $m=5$, while the
smallest dispersions were obtained for the WS networks.  Such
relatively small variations are interesting by themselves as could
be expected that the random choice of the initial node for the walk
would imply paths with markedly different lengths.  However, the
relatively long obtained path lengths, especially for the ER and BA
cases, implied higher probability of taking paths involving the same
nodes.  The average path lengths obtained for the ER networks tended
to be about twice as large as the respective path lengths obtained for
the BA model.  The path lengths resulting for the WS networks were
almost identical to the number of respective nodes in those
structures. The resulting small dispersion of path lengths enhances the
statistical significance of the analysis of measurement changes
described in the following.

First, we consider the ratios of changes in the diameter of the
network, $R(diam)$ (see first group in Table~\ref{tab:res}).  Small
standard deviations were again obtained for all such measurements.  In
the average, the path-star transformation had relatively little effect
in changing the average diameter in the ER networks and BA, but
implied major reduction of diameter in the case of the WS model, an
effect which tended to increase for larger $m$.  The substantial
reductions of diameter obtained for the WS networks are a consequence
of the long initial paths characterizing this case.  Interestingly,
the path-star transformation almost unaffected both the ER and BA
networks.

Regarding the ratio of changes in the clustering coefficient
($R(cc)$), shown in the second group in Table~\ref{tab:res}),
relatively small dispersions were again obtained, except for the ER
networks with $m=3$.  In addition, almost no alteration was again
obtained for the average ratios in the case of the BA networks.
Decreases of about $20\%$ in the clustering coefficient were obtained
for the WS networks.  The changes did not significantly vary with $N$
or $m$. Relatively high increases (all larger than 2) of clustering
coefficient were implied by the path-star transformation in the case
of the ER model.  This is a consequence of the fact that the
concentration of the edges along the path into the middle node tend to
complement the previous connections between non-subsequent nodes in the
path, giving rise to triangles and consequently higher clustering
coefficient.  This effect is less intense for the BA networks because
less edges are typically found between non-subsequent path nodes in
that case (at least for the considered average degrees).

The ratio of changes in the average shortest path lengths are given in
the third group of results in Table~\ref{tab:res}).  Small dispersions
of these ratios were yet again generally observed, enhancing the
statistical significance of the analysis.  While almost no changes in
average path lengths were observed for the ER and BA models,
substantial decreases were implied by the path-star transformation in
the case of the WS networks.  Though the latter result could be
expected (the WS involves an initial long path and large diameter),
the relatively small decrease of average shortest path length obtained
for the ER model was somewhat surprising.  As with the clustering
coefficient, the effect of the path-star transformation tended to
become more accentuated for larger $m$ in the case of the WS networks.

All in all, the above results contained a series of expected and
surprising facts.  Remarkably, relatively small dispersions were
obtained for the initial path length, enhancing the statistical
relevance of our simulations.  Because the BA model involves
relatively short initial paths and is characterized by having several
links connected to hubs, it would be only too natural to expect that
the path-star transformation would have relatively little effect on
that model.  On the other hand, because of the initial sequential
nature of the WS structures, it would be reasonable to expect that the
path-star would imply substantial changes in diameter, clustering
coefficient and average shortest path length for that model.  The
results obtained for the ER contained surprises, especially regarding
the large change of clustering coefficient and small changes of
diameter and average shortest path length observed for those networks.
The small changes in diameter and average shortest path length obtained
for the ER networks are particularly unexpected because the initial
path lengths tended to be twice as larger in the ER as compared to the
BA structures.

\begin{table*}[htb]
\centering
\vspace{1cm}
\begin{tabular}{|c|c||c|c|c|}
  \hline  
Measurement & Configuration&  $ER$         & $BA$          & $WS$ 
          \\ \hline  \hline  
$R(diam)$ & $N=100$, $m=3$ &0.932$\pm$0.005&0.996$\pm$0.100&0.26$\pm$0.060  
          \\  \cline{2-5}
          & $N=100$, $m=5$ &0.880$\pm$0.126&1.003$\pm$0.084&0.399$\pm$0.009
          \\  \cline{2-5}
          & $N=200$, $m=3$ &0.940$\pm$0.085&1.00$\pm$0.00&0.254$\pm$0.161   
          \\  \cline{2-5}
          & $N=200$, $m=5$ &0.964$\pm$0.078&1.00$\pm$0.00&0.34$\pm$0.020   
          \\  \hline \hline
 $R(cc)$  & $N=100$, $m=3$ &2.479$\pm$1.136&1.100$\pm$0.189&0.858$\pm$0.025   
          \\  \cline{2-5}
          & $N=100$, $m=5$ &2.375$\pm$0.353&1.003$\pm$0.084&0.794$\pm$0.010
          \\  \cline{2-5}
          & $N=200$, $m=3$ &2.893$\pm$1.792&1.140$\pm$0.179&0.861$\pm$0.032   
          \\  \cline{2-5}
          & $N=200$, $m=5$ &3.396$\pm$0.599&1.300$\pm$0.179&0.790$\pm$0.006   
          \\  \hline \hline
 $R(spl)$ & $N=100$, $m=3$ &0.916$\pm$0.045&0.996$\pm$0.010&0.480$\pm$0.022
          \\  \cline{2-5}
          & $N=100$, $m=5$ &0.909$\pm$0.028&0.990$\pm$0.012&0.686$\pm0.012$    
          \\  \cline{2-5}
          & $N=200$, $m=3$ &0.912$\pm$0.056&0.996$\pm$0.011&0.416$\pm$0.089    
          \\  \cline{2-5}
          & $N=200$, $m=5$ &0.875$\pm$0.025&0.984$\pm$0.013&0.592$\pm$0.006   
          \\  \hline \hline
 $L$      & $N=100$, $m=3$ &49.72$\pm$15.20&20.78$\pm$9.31&99.58$\pm$2.55 
          \\  \cline{2-5}
          & $N=100$, $m=5$ &81.72$\pm$11.63&45.60$\pm$11.27&100.00$\pm$0.00   
          \\  \cline{2-5}
          & $N=200$, $m=3$ &73.18$\pm$32.56&34.48$\pm$13.76&193.48$\pm$25.96   
          \\  \cline{2-5}
          & $N=200$, $m=5$ &147.76$\pm$14.83&79.64$\pm$18.84&200.00$\pm$0.00 
          \\  \hline
\end{tabular}
\caption{The average $\pm$ standard deviation obtained for the ratios
of change of diameter ($R(diam)$), clustering coefficient ($R(cc)$),
and average shortest path length ($R(spl)$) with respect to each of
the types and configurations of networks. See text for
discussion.}\label{tab:res}
\end{table*}

\subsection{Correlations between Ratios and Initial Path Lengths}

\begin{figure*}
  \vspace{0.3cm} 
  \begin{center}
  \includegraphics[width=0.45\linewidth]{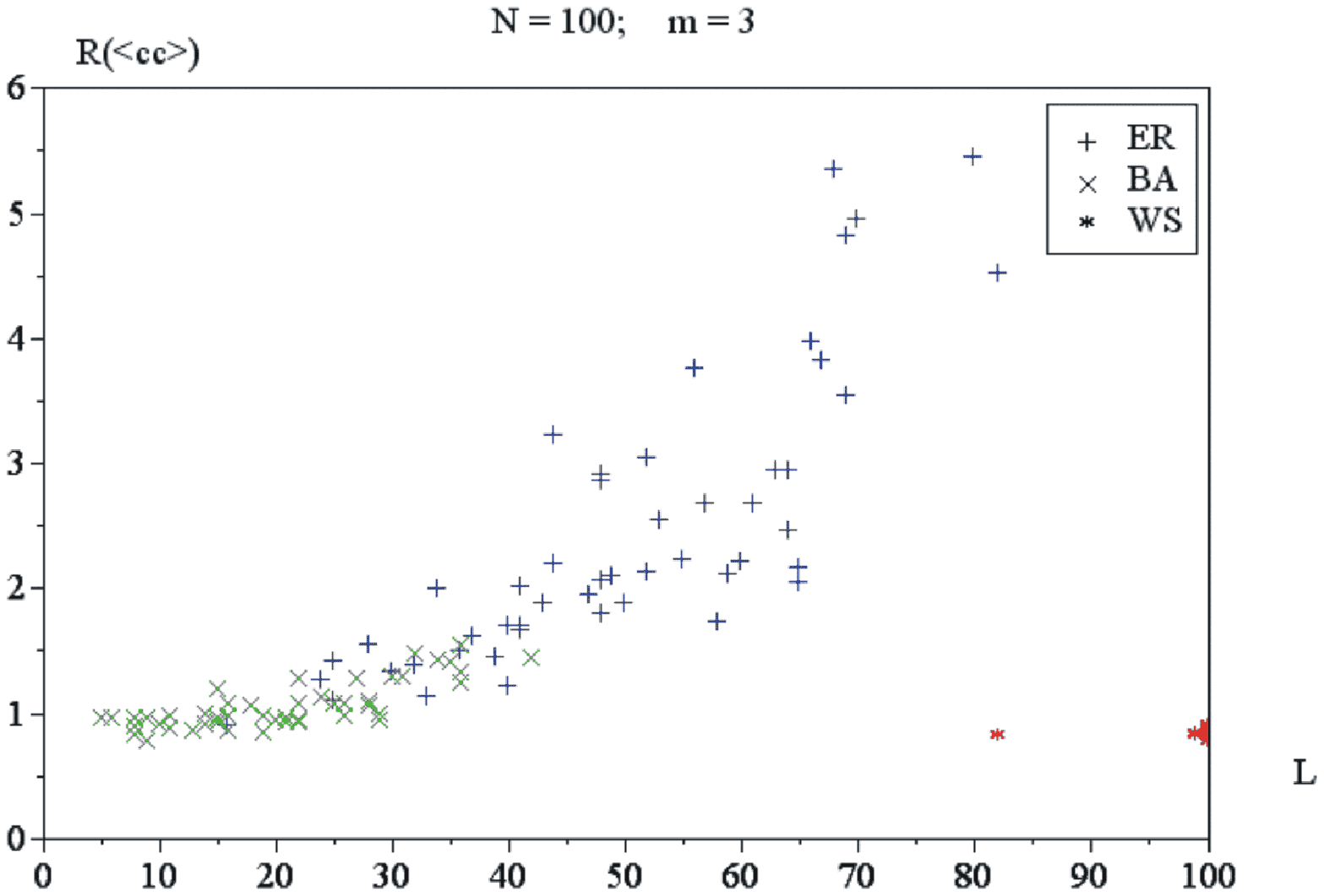}
  \hspace{0.2cm}
  \includegraphics[width=0.45\linewidth]{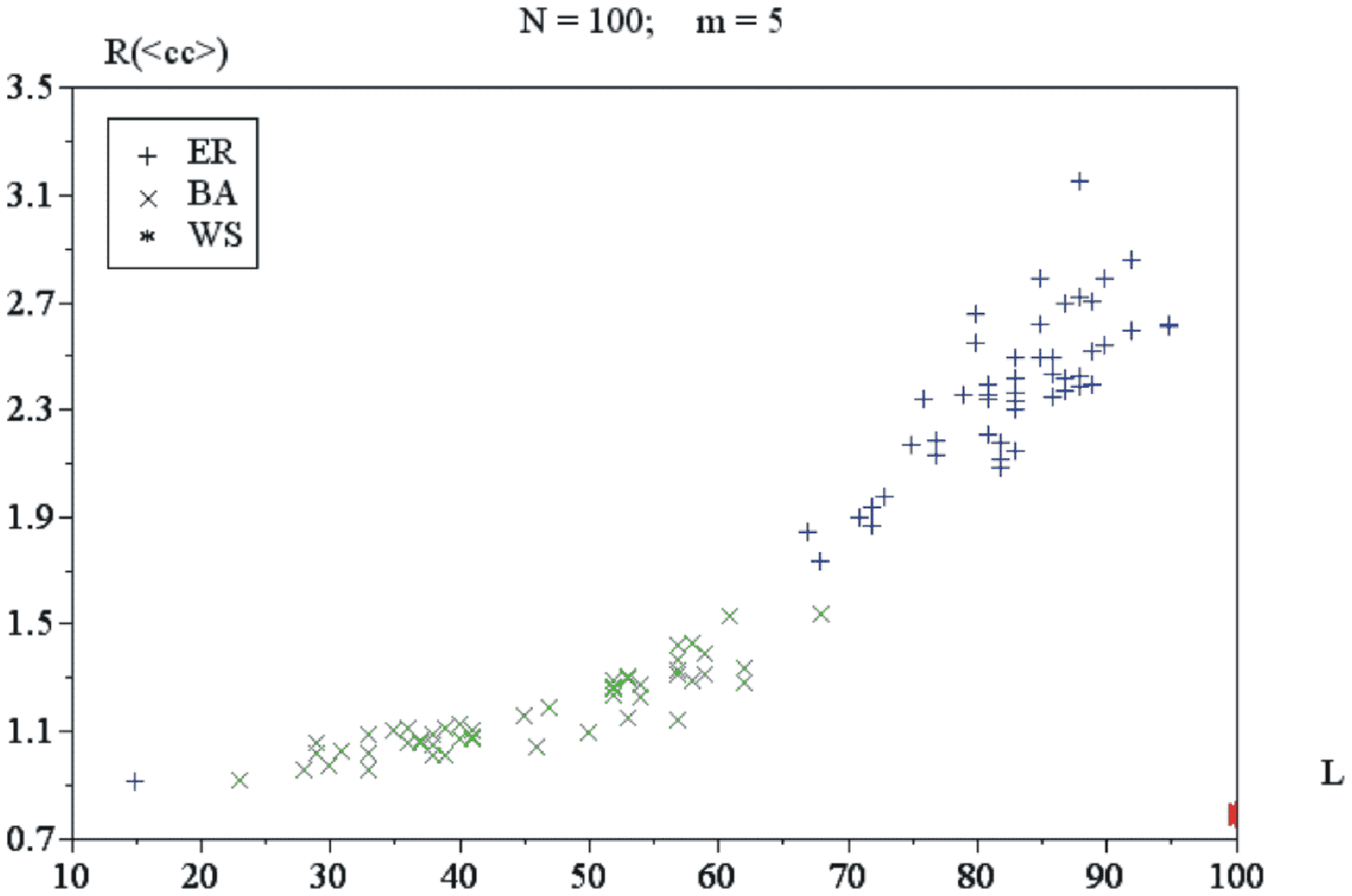} \\
  (a)  \hspace{8cm}  (b) \vspace{0.2cm} \\
  \vspace{0.2cm}
  \includegraphics[width=0.45\linewidth]{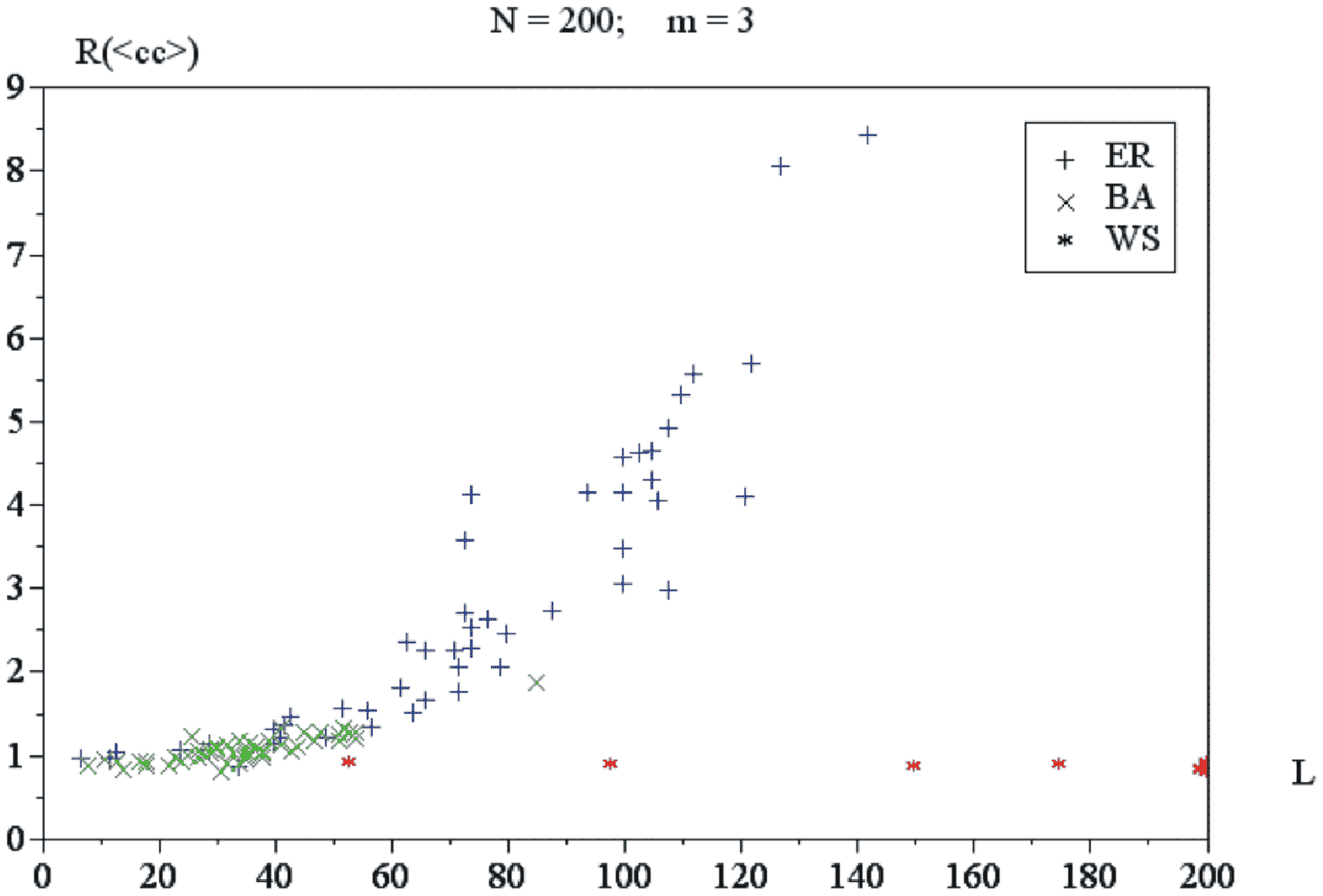}
  \includegraphics[width=0.45\linewidth]{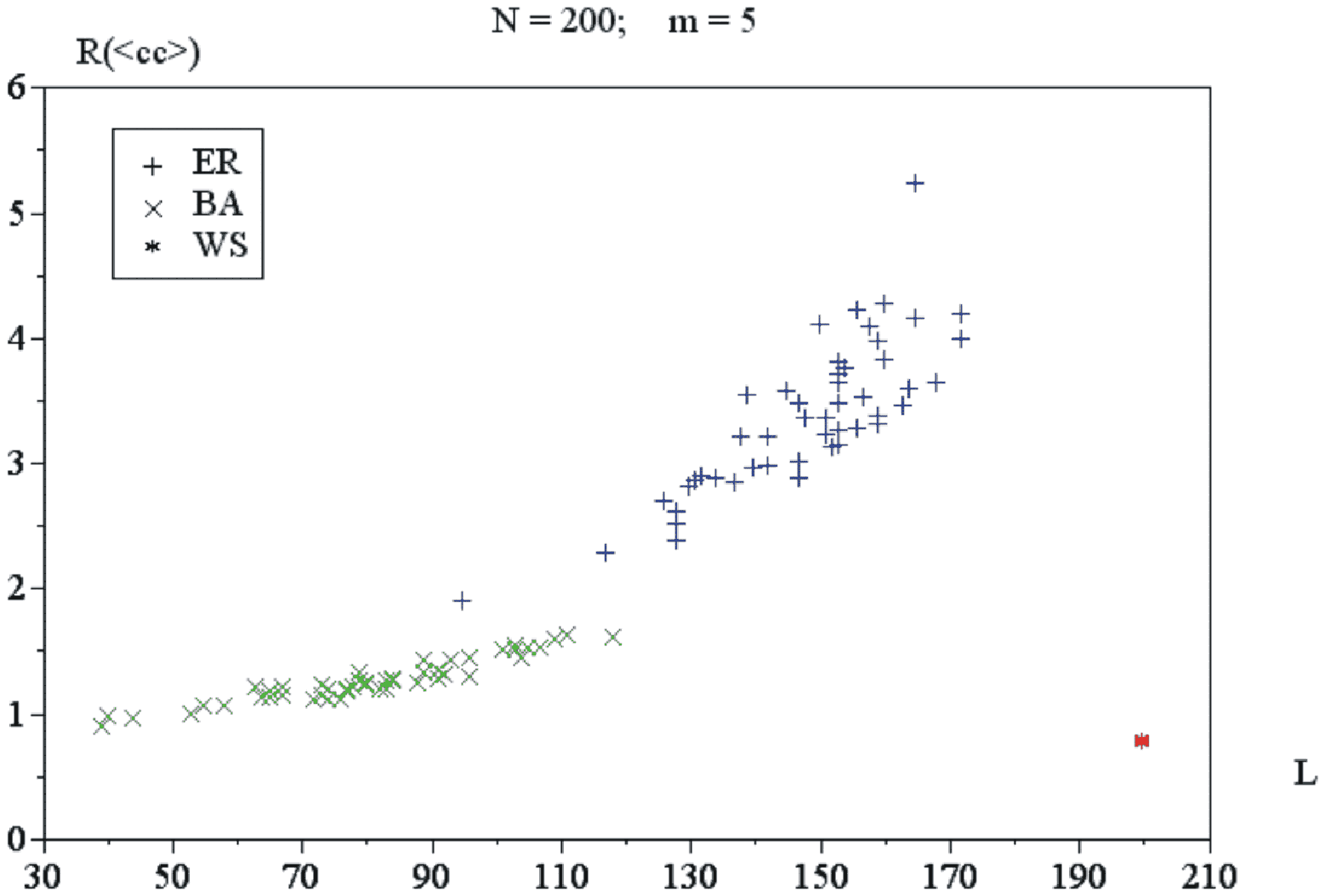} \\
  (c)  \hspace{8cm}  (d)  \\
   \caption{The scatterplots obtained by plotting the ratio of changes
   in the clustering coefficient, $R(cc)$, in terms of $L$ for the
   four considered configurations, i.e.: $N=100$, $m=3$ (a); $N=100$,
   $m=5$ (b); $N=200$, $m=3$ (c) and $N=200$, $m=5$ (d).  This
   scatterplot indicates a well-defined tendency of $R(cc)$ to
   increase with $L$ in the case of the ER and BA networks, while
   almost identical ratios were obtained for the WS model.}
   \label{fig:cc}
  \end{center}
\end{figure*}

\begin{figure*}
  \vspace{0.3cm}
  \begin{center}
    \includegraphics[width=0.45\linewidth]{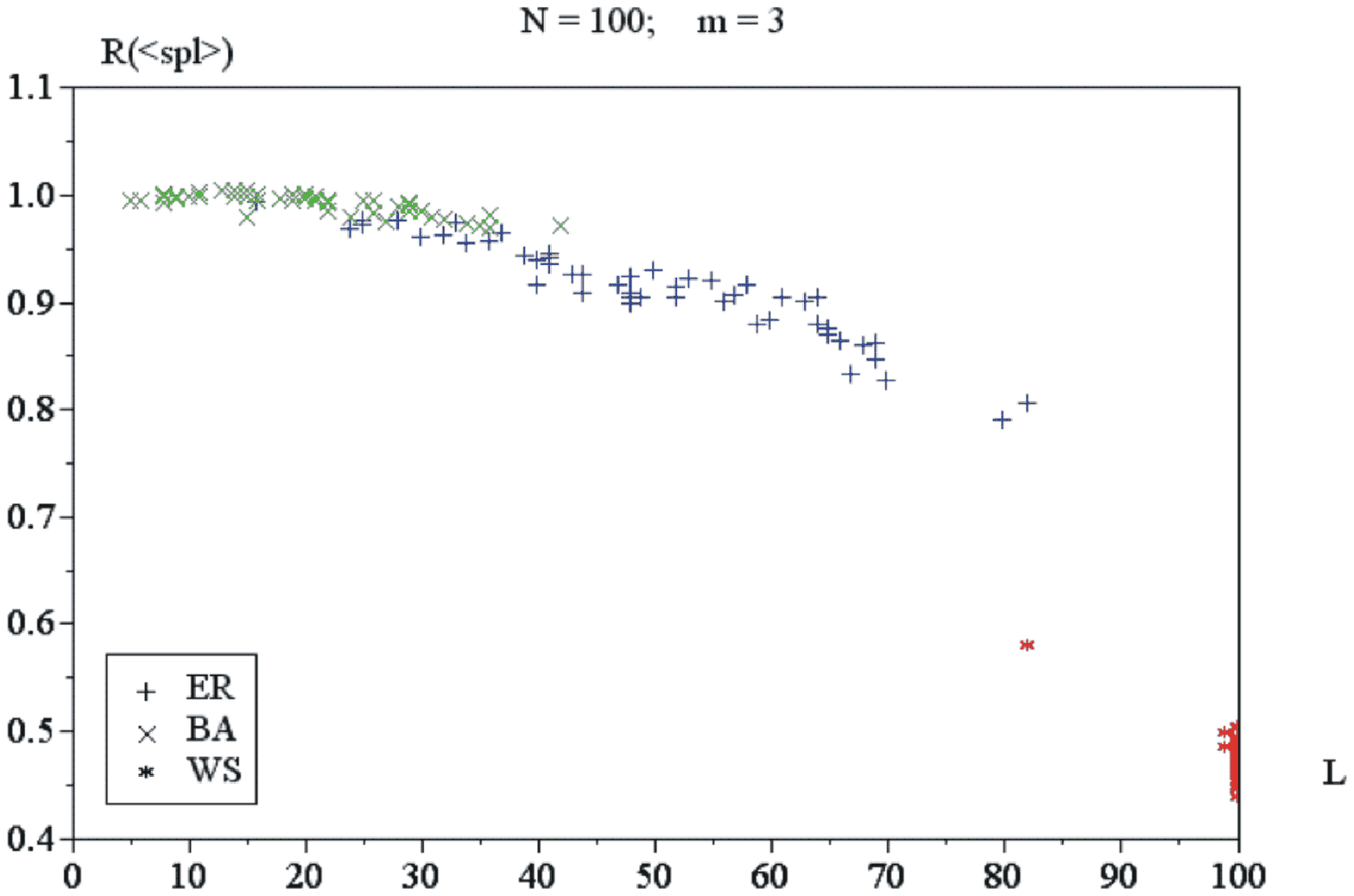}
    \hspace{0.2cm}
    \includegraphics[width=0.45\linewidth]{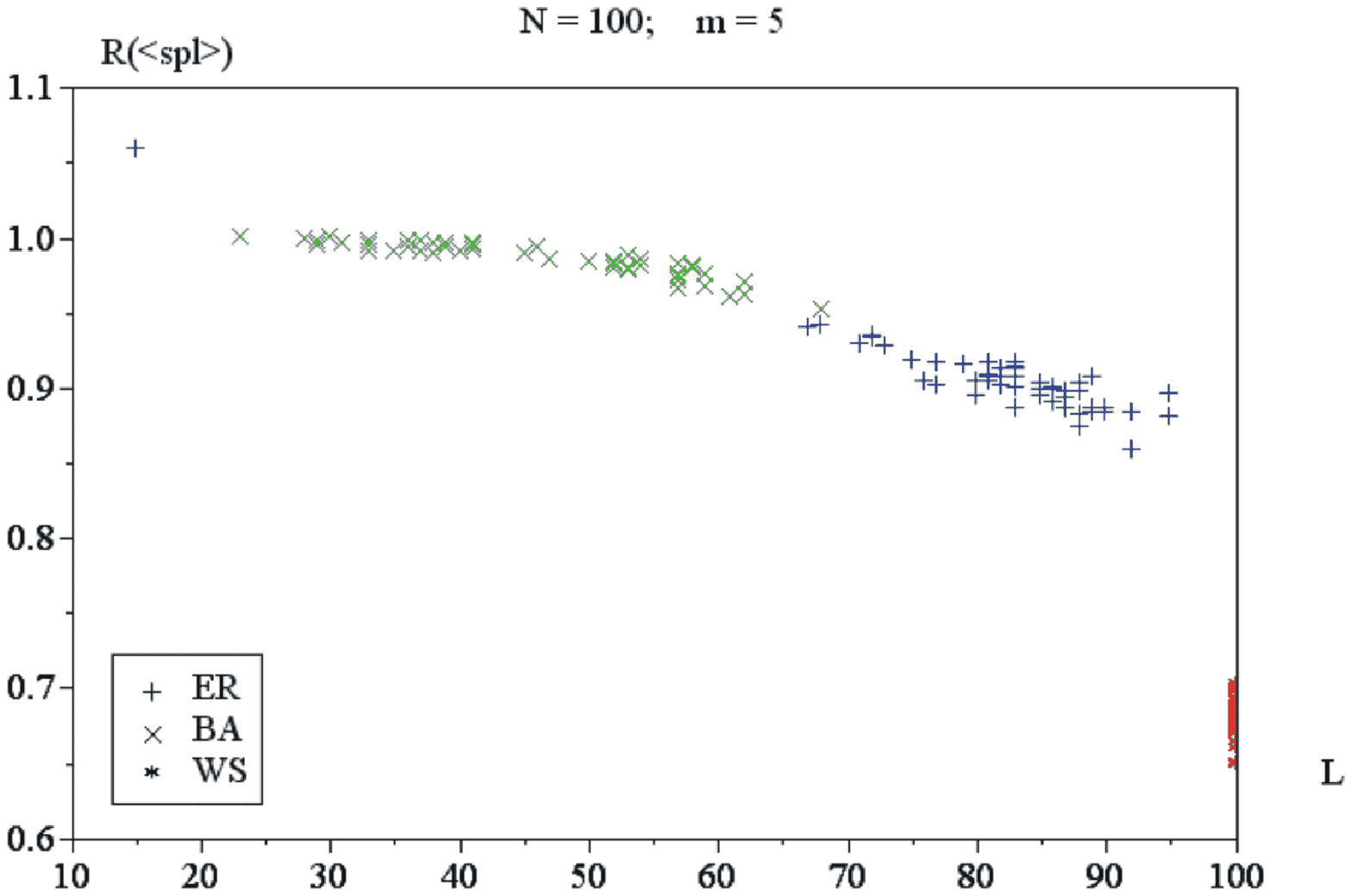} \\
    (a)  \hspace{8cm}  (b) \vspace{0.2cm} \\
    \vspace{0.2cm}
    \includegraphics[width=0.45\linewidth]{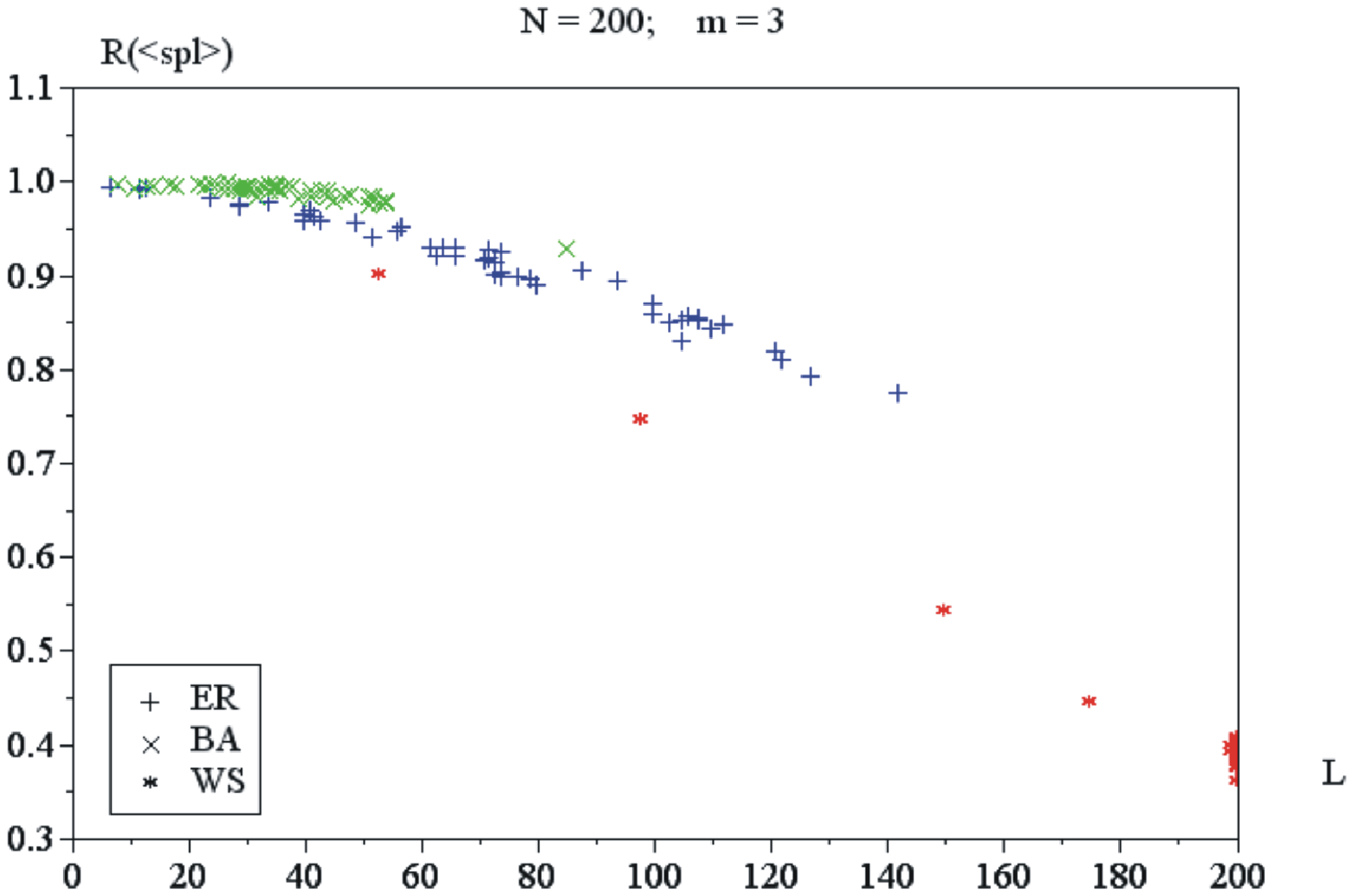}
    \hspace{0.2cm}
    \includegraphics[width=0.45\linewidth]{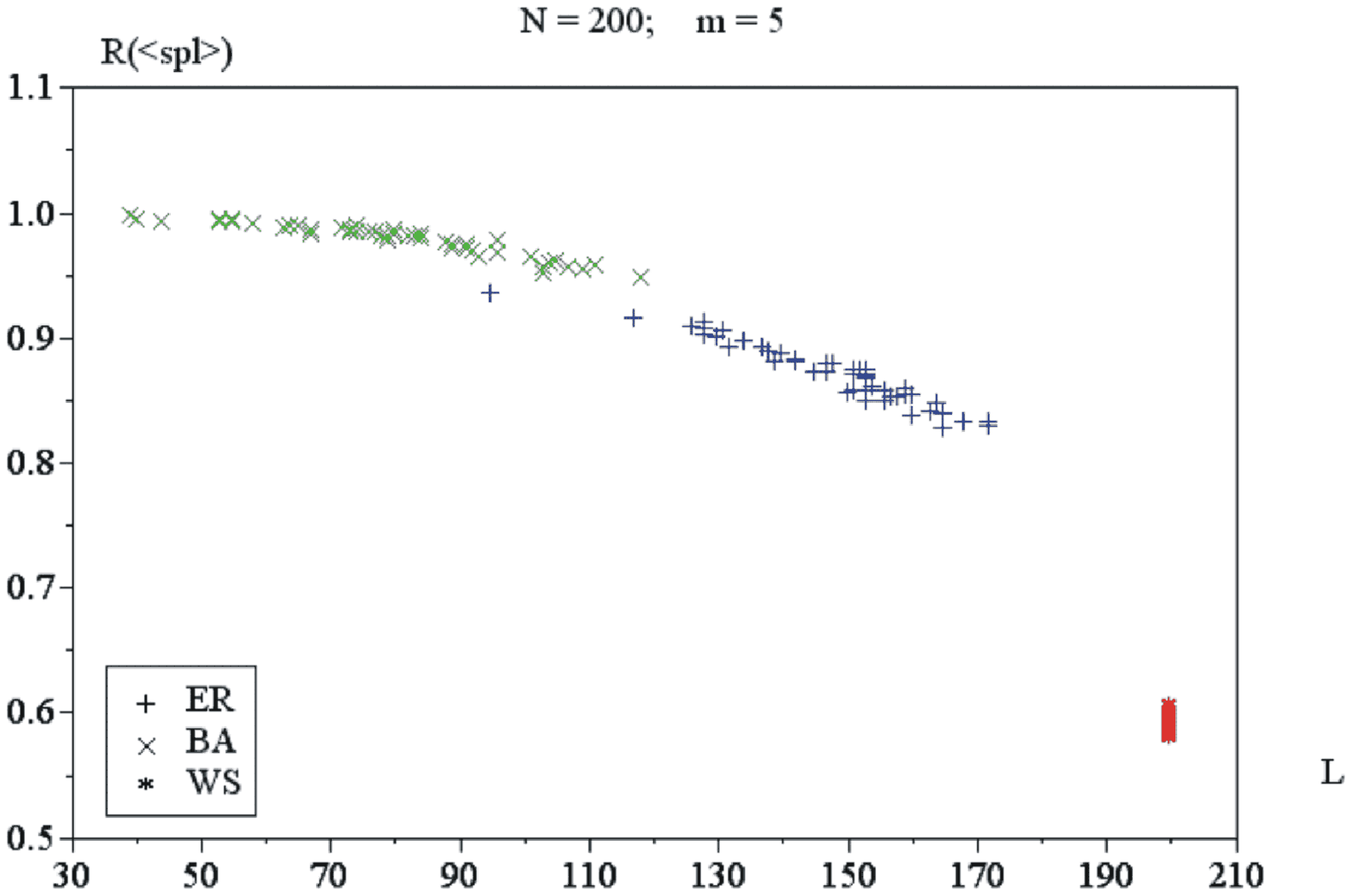} \\
    (c)  \hspace{8cm}  (d) \\
  \caption{The scatterplots obtained by plotting the ratio of change
  in the average shortest path length, $R(spl)$, in terms of $L$ for
  the four considered configurations, i.e.: $N=100$, $m=3$ (a);
  $N=100$, $m=5$ (b); $N=200$, $m=3$ (c) and $N=200$, $m=5$ (d). These
  results indicate a clear tendency of $R(spl)$ to decrease with $L$
  in the cases of ER and BA networks.  However, very little variation
  of $R(spl)$ was observed in the case of the WS model.}
  \label{fig:spl}
  \end{center}
\end{figure*}

Despite the relatively small standard deviation of the initial path
lengths obtained for the considered models, it is interesting to
investigate the relationships between those values and the three
considered ratios $R(diam)$, $R(cc)$, $R(spl)$.  Because the diameters
involve only a few integer values (e.g. 2, 3, etc.), such correlations
are not considered here.  The scatterplots obtained by plotting
$R(cc)$ and $(spl)$ against $L$ are presented in Figures~\ref{fig:cc}
and ~\ref{fig:spl}, respectively.  Each of these two figures includes
all the 50 realizations obtained for each of the configurations of the
three considered network models.

As can be appreciated from Figure~\ref{fig:cc}, the ratio of changes
in the clustering coefficient, as implied by the path-star
transformation, tended to increase substantially with $L$ in the case
of the ER and BA models, while very similar (and small) ratios were
obtained for the WS networks.  Interestingly, the increase for ER and
BA is more accentuated for $m=3$ than for $m=5$.  Observe also that
the dispersion of the ratios $R(cc)$ obtained for the ER and BA models
tended to increase substantially with $L$.  This effect is
particularly counter-intuitive in the sense that the path-star
transformations performed on particularly long initial paths
(especially in the ER case) --- which therefore encompass most of the
network nodes --- could be expected to lead to paths containing
similar nodes and, therefore, similar connectivity.

Figure~\ref{fig:spl} shows the scatterplot obtained by plotting the
ratios of change in the average shortest path length, i.e. $R(spl)$,
in terms of $L$. The path-star transformation tends to imply a
reduction of the average shortest path (i.e. $R(spl)<1$) in all cases,
which more intense effects for larger values of the initial path
length $L$.  Unlike what was found for the clustering coefficient,
relatively small dispersions of $R(spl)$ are observed for each value
of $L$.  The behavior of $R(spl)$ is completely different from those
obtained for the ER and BA cases.  More specifically, the reductions
of $R(spl)$ observed for the WS networks are much more accentuated and
evolve in a distinct fashion when $L$ increases.  Actually, an almost
linear decrease of $R(spl)$ can be identified in
Figure~\ref{fig:spl}(c).

\section{Concluding Remarks}

Two of the most important connectivity patterns characterizing complex
networks are paths and hubs, which have a dual nature.  The present
work has investigated how the transformation of paths into stars (hubs
in the case of long paths) in networks tend to change the properties
of ER, BA and WS network as far as the properties of diameter,
clustering coefficient and average shortest path length are
concerned. For each considered network, a path was randomly chosen and
transformed into a star.  Ratios between the above measurements
obtained after and before the transformation were calculated in order
to allow the identification of the major effects implied by the
path-star transformation with respect to each of the three types of
networks.  A series of interesting results were obtained which provide
insights not only about the effects of the transformation but also on
the inherent structure of the connectivity in each type of the models.
The first interesting result is that the length of the randomly chosen
path tended to present relatively small to moderate dispersion,
enhancing the statistical relevance of the analysis performed for the
ratios.  Because of its intrinsic more sequential nature (starts as a
cycle, a fact reflected in the obtained very high values of $L$), the
WS was the network model more intensely affected by the path-star
transformation.  Contrariwise, very little changes were observed for
the BA model, which involves a significant number of hubs
concentrating the edges into stars and reducing the average shortest
path length.  Interestingly, the ER model responded in surprising ways
to the path-star transformation.  While the diameter and average
shortest path lengths underwent relatively small changes after the
path-star transformation, a substantial increase of average clustering
coefficient was observed for all configurations.  This effect is
explained by the tendency of the path-star transformation to complete
triangles involving edges between non-subsequent nodes along the
transformed path.  The correlations between the ratios of change and
the length of the transformed path were also analyzed, confirming the
markedly distinct response of the WS networks to the transformation.
The effect of the path-star transformation on the average clustering
coefficient was also found to become more pronounced for smaller
values of $m$ (i.e. average degree).  By quantifying the responses of
the three categories of networks to the path-star transformation, the
reported investigation allows this type of network modification to be
applied in order to change (hopefully improve) specific features of
given real-world networks.

Several are the possible future developments motivated by this work.
To begin with, motivated by the relationship established between long
paths and hubs (dual), it would be interesting to invest additional
efforts in characterizing the longest path typically found in
different types of networks.  Such a study would have immediate
implications for the susceptibility of the models to the path-star
transformation.  Second, it would be natural to consider more than one
subsequent path-star transformations.  Other interesting future works
include the consideration of real-world networks, the inverse
star-path transformation, and the application of additional
measurements (especially concentric
features~\cite{Costa_PRE:2004,Costa_Andrade:2007}).

\begin{acknowledgments}
Luciano da F. Costa thanks CNPq (308231/03-1) and FAPESP (05/00587-5)
for sponsorship.
\end{acknowledgments}

\bibliography{path_star}
\end{document}